\documentclass[twocolumn,showpacs,preprintnumbers,amsmath,amssymb,nofootinbib]{revtex4}
\usepackage{tabularx,graphicx}\begin{document}
\newcommand{\beq}{\begin{equation}}
\newcommand{\eeq}{\end{equation}}
\newcommand{\beqn}{\begin{eqnarray}}
\newcommand{\eeqn}{\end{eqnarray}}
\newcommand{\bmath}{\begin{subequations}}
\newcommand{\emath}{\end{subequations}}
\title{Does the h-index have predictive power?}
\author{J. E. Hirsch }
\address{Department of Physics, University of California, San Diego\\
La Jolla, CA 92093-0319}

\begin{abstract} 
Bibliometric measures of individual scientific achievement are of particular interest if they can
be used to predict future achievement. Here we report results of an empirical study
of the predictive power of the h-index compared to other indicators. Our findings
indicate that the h-index is better than  other indicators considered (total citation count, citations per paper, and total paper count) in predicting
future scientific achievement. We discuss  reasons for the superiority of the h-index.
\end{abstract}
\maketitle 

\section{introduction}
The h-index of a researcher is the number of papers coauthored by the researcher with at least h citations each\cite{hindex}. We have recently  proposed it as a
representative measure of individual scientific achievement. Other commonly used bibliometric measures of individual scientific achievement are total number of
papers published ($N_p$) and total number of citations garnered ($N_c$). Recently, Lehmann et al have argued\cite{lehmann1,lehmann2} that the mean number of citations per paper 
($n_c=N_c/N_p$) is a superior indicator. Here we wish to address the question: which of these four measures is best able to predict future scientific achievement?

For the purposes of this paper we do not wish to dwell on the controversial question of what is the optimal definition of scientific achievement\footnote{Philip Ball, "Is it any good? Measuring scientific merit", presentation at Deutsche Physikalische Gesellschaft meeting, Regensburg, Germany, March 26-30, 
2007, http://www.agrfoto.com/philipball/docs/pdf/anygood.pdf.}. 
We are not interested in measuring the $past$ achievement of an individual e.g. for the purpose of awarding a prize or for  election to a prestigious academy, 
but rather in predicting $future$ achievement. So we could simply bypass this question by defining scientific achievement by the bibliometric 
measure under consideration, and ask: which measure is better able to predict its future values? For example, how likely is a researcher that today
has a large number of citations to gain a large number of citations in future years? To the extent that a bibliometric measure reflects particular traits of the
researcher rather than random events, it should have higher predictive power than another measure that is more dependent of random events.
For example, we argued in ref. 1 that the total number of citations $N_c$ "may be inflated by a small number of 'big hits',
which may not be representative of the individual if he/she is coauthor with
many others on those papers". For that individual, the present $N_c$ value is not likely to be a good predictor of his/her future $N_c$ values.

Alternatively, among the indicators listed in the first paragraph, it may be argued that the total number of citations $N_c$ is the one that best reflects scientific achievement,
since it gives a measure of the integrated impact of a scientist's work on the work of others. Then we would like to know: which indicator is best able
to predict $N_c$ at a future time? It is certainly not obvious that the answer is $N_c$ itself.

There are two slightly different questions of interest: (a) Given the value of an indicator at time $t_1$, $I(t_1)$, how well does it predict the value
of itself or of another indicator at a future time $t_2$, $I'(t_2)$? This question is of interest for example in trying to decide between different
candidates for a faculty position. A possible consideration might be: how likely is each candidate to become a member of the National Academy of Sciences
20 years down the line? For that purpose, one would like to rank the candidates by their expected $cumulative$ achievement 
after 20 years. Which means in particular that citations obtained $after$ time $t_1$ to papers written $before$ time $t_1$ are relevant. 
(b) Instead, to award grant money or other resources for future research at time $t_1$ one would like to rank the candidates by their expected
scientific output $from$ $time$ $t_1$ $on$ to some future time. In deciding who should get a grant it should be
irrelevant how many more citations the earlier papers of that individual are expected to collect in future years.

\section{procedure}
We use the ISI Web of Science database in the 'general search' 
mode.\footnote{In using the very valuable ISI resource for individual evaluations, one should keep in mind that it has limitations, e.g.: (i) it will of course miss
citations where the author's name is misspelled; (ii) books, book chapters and most conference proceedings are not 
included; (iii)  citations to "Rapid Communications" papers in Physical Review that include (R) in the citation are currently not counted by ISI.} 
ISI has recently incorporated tools under 'Author Finder' that help to discriminate
between different researchers with the same name. Once the publications of a researcher are identified, ISI provides in the 'Citation Report' the total
number of citations $N_c$, total number of papers $N_p$, citations per paper $n_c=N_c/N_p$ and the h-index, all 
$at$ $the$ $present$ $time$.

ISI also allows one to restrict the time frame of the papers' publication date, so it is easy to find $N_p(t_1)$, the number of papers published up to
year $t_1$, or $N_p(t_1,t_2)$, the number of paper published between times $t_1$ and $t_2$. 
However, ISI does not provide a similarly simple way to obtain these values for the other indicators under consideration. To obtain this information, one needs to export ('save') the information in the Citation Report into an Excel file and add up the citations in the time interval desired. It is a straightforward but tedious
procedure.

For illustration we show in Fig. 1 the h-index and the number of citations as function of time for two prominent physicists, a theorist and an
experimentalist. As we conjectured in Ref. 1, the h-index follows approximately a linear behavior with time and the total number of citations is
approximately quadratic with time. Similar behavior is found in many other cases.

\begin{figure}
\resizebox{6.5cm}{!}{\includegraphics[width=8cm]{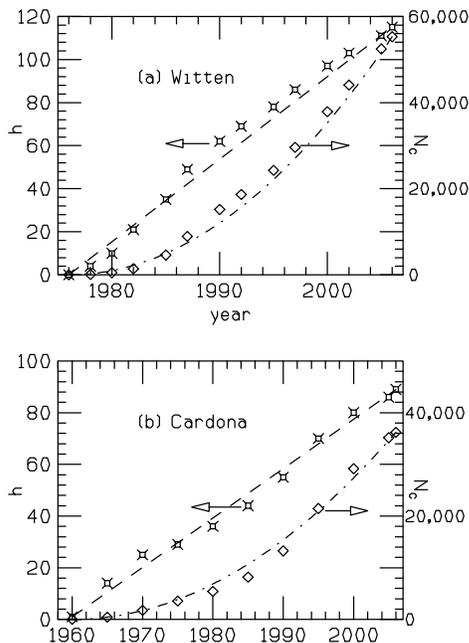}}
\caption{h-index versus time (left scale) and total number of citations ($N_c$) versus time (right scale) for E. Witten (theorist) and M. Cardona (experimentalist).
The dashed and dash-dotted lines show linear and quadratic fits to h-index and $N_c$ respectively.}
 \label{Fig. 2}
\end{figure}

\section{Sample PRB80}
Because citation patterns vary between fields and also between subfields and there are also trends with time, 
we chose to look at authors in a single subfield and of comparable
scientific age to compare the
predictive power of the various indicators. Ideally we would like to pick a random subset of all physicists that earned their Ph.D. in a given subfield in a given year
and published in that subfield throughout their career.
However we have no practical way to make such a selection. As alternative, 
we picked a sample of 50 physicists that started publishing around 1980 by the following procedure:

\begin{table}
\caption{Averages and standard deviations in different time frames for the four indicators considered for the sample PRB80. 
The last column gives $a=N_c/h^2$.}
\begin{tabular}{l || c | c  | c  | c  |c}
Time frame & h, $\sigma_h$ & $N_c$, $\sigma_{N_c}$ & $N_p$, $\sigma_{N_p}$ & $n_c$, $\sigma_{n_c}$& $a$, $\sigma_a$  \cr
\tableline
yr 1-12 & 14, 6 & 740, 720 &  45, 26  & 16, 10 &3.5, 1.2  \cr
\tableline
yr 1-24 & 24, 12 & 2452, 2471 &  90, 51  & 25, 14 &4.1, 1.7  \cr
\tableline
yr 13-24 & 12, 8 & 759, 831 &  45, 33  & 13, 10 &3.8, 1.7  \cr
\tableline
\end{tabular}
\end{table}

(i) We considered papers published in Physical Review B in 1985 that have today citations in the range 45 to 60 (an arbitrary choice, simply
to avoid extremes). In practice we started with papers with 60 citations and went as far down as needed to get the number of authors desired for our sample. 

(ii) From the authors of those papers, we selected those that had published their first paper between 1978 and 1982.
 
Because of the journal used for the selection (Phys.Rev.B) the sample contained mostly physicists that published in the field of condensed matter 
physics throughout their career. There was however a small subset of the sample that subsequently switched to other subfields.

We then looked at the publication records of these authors during the first 12 years of their career (starting with their first published paper) and
in the subsequent 12 years. In table I, we show the average and standard deviation values of the four indicators considered in the first 12 years,
first 24 years, and years 13 to 24, all measured from the publication year of the first paper. It can be seen that $h$, $N_p$ and
$n_c$ increase by approximately a factor of 2 in comparing 12 year and 24 year periods and $N_c$ by approximately a factor of 4, as 
expected. The last column in the table shows the average and standard deviation of $a=N_c/h^2$ for this sample.  In Ref. 1 $a$ was observed to be 
typically in the range $3$ to $5$.

As discussed earlier we would like to know:

(a) How well does the performance during the first 12 years predict the cumulative achievement over the entire 24 year period?

(b) How well does the performance during the first 12 years predict the performance in the subsequent 12 years?

Because the number of citations is expected to grow quadratically with time, we used $\sqrt{N_c}$ as measure of total citations.

\begin{figure}
\resizebox{9.0cm}{!}{\includegraphics[width=8cm]{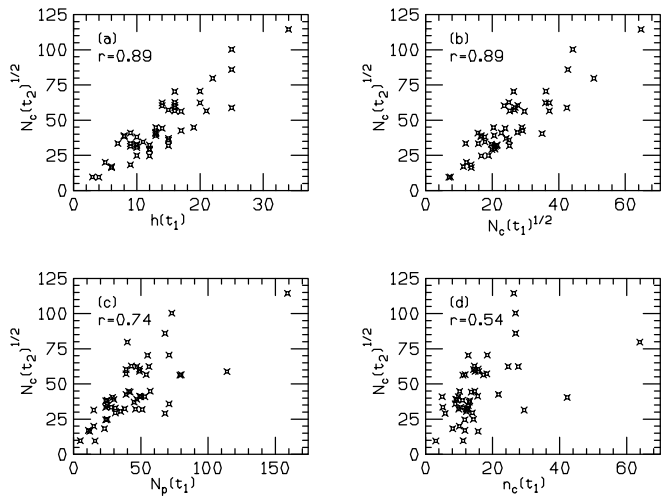}}
\caption{Scatter plot of total number of citations $N_c$  after $t_2=$24 years versus the value of the various indicators at $t_1$=12 years (t measured from the date of the
first publication). $N_c=$number of citations, h=h-index, $N_p=$number of papers, $n_c$=mean number of citations per paper. r is the correlation
coefficient, r=1 corresponding to perfect linear correlation.}
 \label{Fig. 2}
\end{figure}

First we consider the predictive power of the various indicators after the first 12 years ($t_1$) for the cumulative achievement in the 24-year period ($t_2$). In Fig. 2 we show the  total number of
citations after 24 years versus each indicator after 12 years for each member of the sample, and their 
correlation coefficient $r$ (=covariance / product of standard deviations).
 It can be seen that the h-index and the number of
citations $N_c$  at time $t_1$ are the best predictors of cumulative citations at the future time $t_2$, with correlation coefficient $r=0.89$. The number of papers correlates somewhat less
($r=0.74$), and the number of citations per paper $n_c$  has lowest correlation with cumulative citations, with $r=0.54$.

According to these results, if one wishes to select among various candidates at time $t_1$ the one(s) that will have the largest number of
citations at the later time $t_2$, the h-index or the number of citations at time $t_1$ are   good selection criteria. A candidate with low h or low $N_c$
at time $t_1$ will not have a high $N_c$ at time $t_2$. Instead, a candidate with low $N_p$ or low $n_c$ at time $t_1$ has a much higher chance
of ending up with high $N_c$ at time $t_2$.

  \begin{figure}
\resizebox{9.0cm}{!}{\includegraphics[width=8cm]{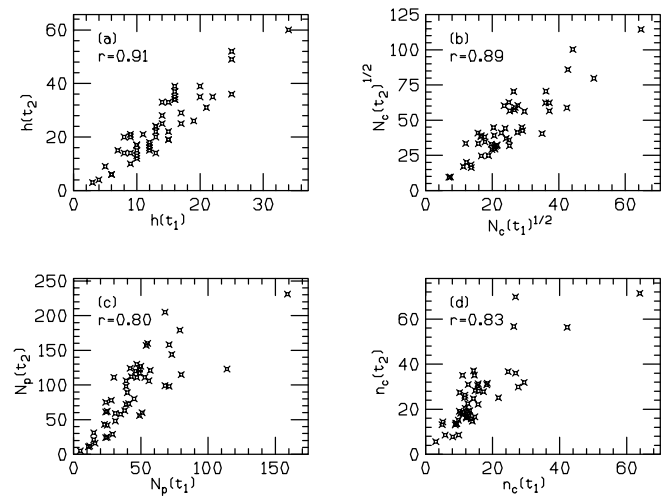}}
\caption{Predictive power of each indicator at time $t_1$=12 years for the value of the same indicator at time $t_2$=24 years for sample PRB80.}
 \label{Fig. 2}
\end{figure}

Figure 3 shows the ability of each indicator to predict its own cumulative value. Here, the differences between indicators is smaller and 
the correlation coefficient is high in all cases. Still, the h-index shows the largest predictive power, with $r=0.91$. 
I.e., a researcher with high h-index after 12 years is highly likely to have a high h-index after 24 years. 

It is more difficult for the indicators at time $t_1$ to predict scientific achievement  occurring only  in the subsequent period, i.e. without taking into account the citations after time $t_1$ to work performed prior to $t_1$. As discussed, one would like to make such predictions to
decide on allocation of research resources. In figure 4, the ability of the indicators at time $t_1$ to predict citations to papers written in the 
$t_1-t_2$ time interval is considered. The highest correlation coefficient occurs for the h-index ($r=0.60$), and the lowest for mean number of citations
per paper ($r=0.21$). Similarly, as shown in Fig. 5, the ability of each index to predict itself is largest for the h-index ($r=0.61$) and lowest for number of citations
per paper ($r=0.23$).

So if we choose to measure 'scientific achievement' by either total citation count $N_c$ or by the h-index, these results imply that
(at least in this example) the h-index has the highest ability to predict $future$ scientific achievement. 
In fact, even choosing the number of papers $N_p$ as the measure of achievement the h-index yields the highest predictive power, as shown in Fig. 6: $r=0.49$, versus
$r=0.43$, $r=0.42$ and $r=0.092$ for $N_p$, $N_c$ and $n_c$ as predictors respectively.
In allocating research resources (eg grant funding) 
to otherwise comparable researchers,   if the goal
is to maximize the expected return on the investment as measured by either $N_c$ or h-index or $N_p$, 
we suggest that these results should be considered. 
If one chose instead to use as indicator of scientific achievement 
the mean number of citations per paper (following Lehmann et al\cite{lehmann1,lehmann2}) our results suggest that 
(as in the stock market) 'past performance is not predictive of future performance'.

  \begin{figure}
\resizebox{9.0cm}{!}{\includegraphics[width=8cm]{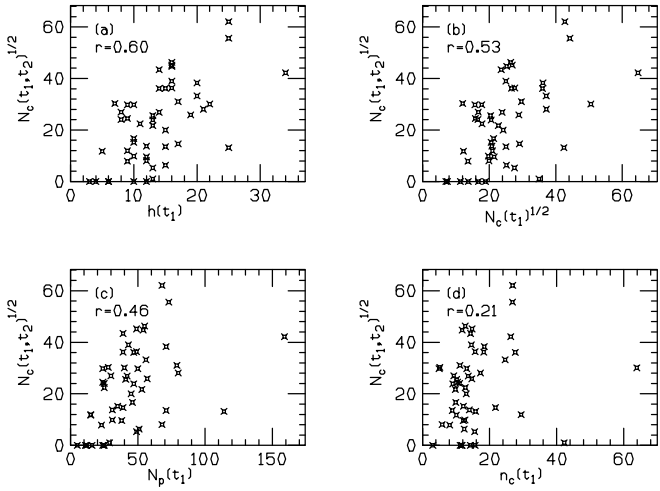}}
\caption{Predictive power of each indicator at time $t_1$=12 years for the number of citations to papers published in the $t_1-t_2$ time interval, with  $t_2$=24 years, for sample PRB80.}
 \label{Fig. 2}
\end{figure}

 \begin{figure}
\resizebox{9.0cm}{!}{\includegraphics[width=8cm]{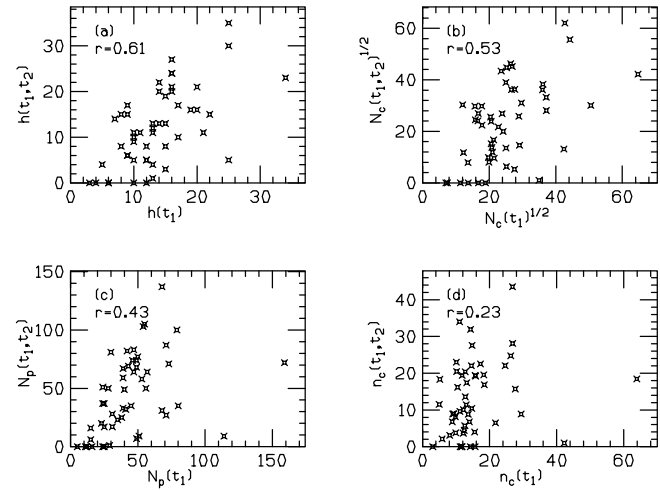}}
\caption{Predictive power of each indicator at time $t_1$=12 years for the value of the same indicator for the papers published
 in the $t_1-t_2$ time interval, with  $t_2$=24 years, for sample PRB80.}
 \label{Fig. 2}
\end{figure}

  \begin{figure}
\resizebox{9.0cm}{!}{\includegraphics[width=8cm]{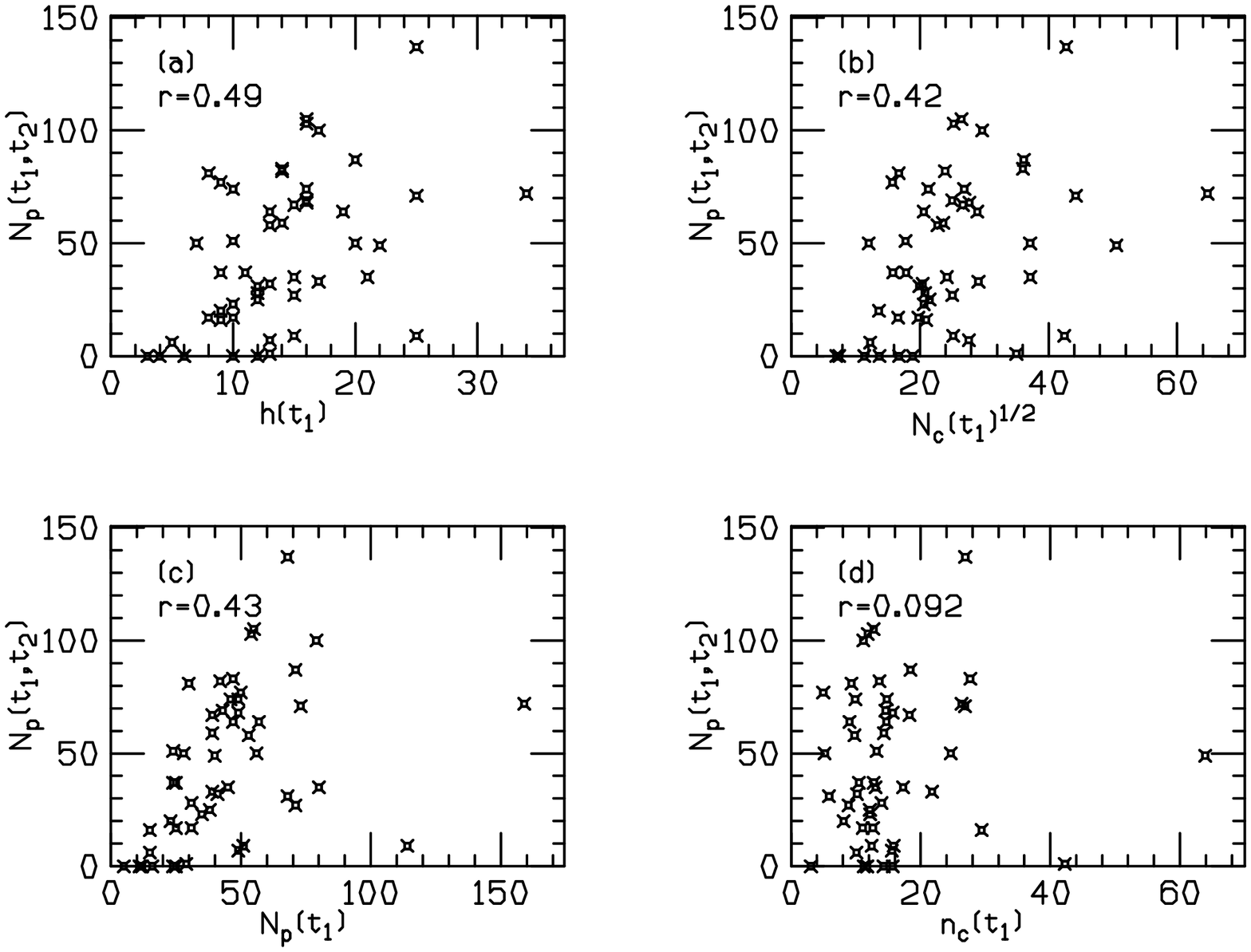}}
\caption{Predictive power of each indicator at time $t_1$=12 years for the number  papers published in the $t_1-t_2$ time interval, with  $t_2$=24 years, for sample PRB80.}
 \label{Fig. 2}
\end{figure}

\section{Sample APS95}

As a second example, we consider the set of physicists elected to fellowship in 1995 by the Division of Condensed Matter Physics of the American Physical Society. (The list is available
online at $http://dcmp.bc.edu/page.php?name=fellows\_95$). From the list of 29 individuals, two were excluded   because it was difficult to identify their publications  due to name overlaps. 
We evaluated the indicators for this group up to the year 1994 (right before being elected to fellowship), up to 2006, and in the 12  years  from 1995 to 2006. The averages
and standard deviations are shown in Table II.

\begin{table}
\caption{Averages and standard deviations in different time frames for the four indicators considered for the sample APS95. 
The last column gives $a=N_c/h^2$.}
\begin{tabular}{l || c | c  | c  | c  |c}
Time frame & h, $\sigma_h$ & $N_c$, $\sigma_{N_c}$ & $N_p$, $\sigma_{N_p}$ & $n_c$, $\sigma_{n_c}$& $a$, $\sigma_a$  \cr
\tableline
up to yr 94 & 22, 7 & 1887, 1348 & 114, 53 & 17, 7 & 3.6, 0.9  \cr
\tableline
up to yr 06 & 36, 13 & 5741, 4468 & 205, 122 & 32, 41 & 4.2, 1.6  \cr
\tableline
yrs 95-06 & 19, 11 & 1718, 1760 & 91, 87 & 19, 12 & 4.1, 2.1  \cr
\tableline
\end{tabular}
\end{table}
 
 Figure 7 shows the number of citations in the 12 years after being elected to fellowship versus each of the indicators up to the year 1994. The correlations here are weaker
 than in the first example, nevertheless the h-index shows a stronger correlation (r=0.49)  that all other indicators. Similarly, Figure 8 shows that 
 the h-index is a better predictor of itself (r=0.54)  than any of the other indicators.
 
 Incidentally, note the large dispersion in the values of the indicators at time $t_1$ (e.g. $h$ ranging from $9$ to $43$, $N_c$ from $482$ to $7471$, 
 $N_p$ from $19$ to $248$), 
 which indicates that the APS fellowship committee does
 not rely (for better or for worse) on any of these numerical indicators as a deciding factor for election to fellowship.
 
 The data for cumulative achievement up to 2006 are shown in Figs. 9 and 10. It can be seen that the pattern is similar to Figs. 2 and 3, the corresponding graphs for sample PRB80.

  \begin{figure}
\resizebox{9.0cm}{!}{\includegraphics[width=8cm]{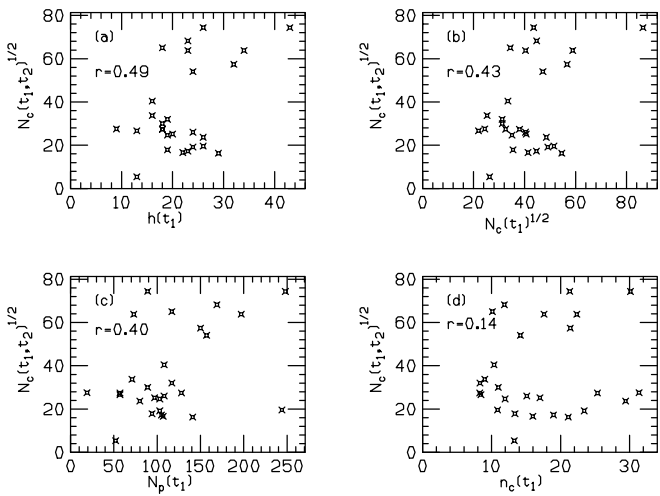}}
\caption{Predictive power of each indicator at year 1994 for the number of citations to papers published in the $1995-2006$ time interval for the sample APS95.}
 \label{Fig. 2}
\end{figure}
 
  \begin{figure}
\resizebox{9.0cm}{!}{\includegraphics[width=8cm]{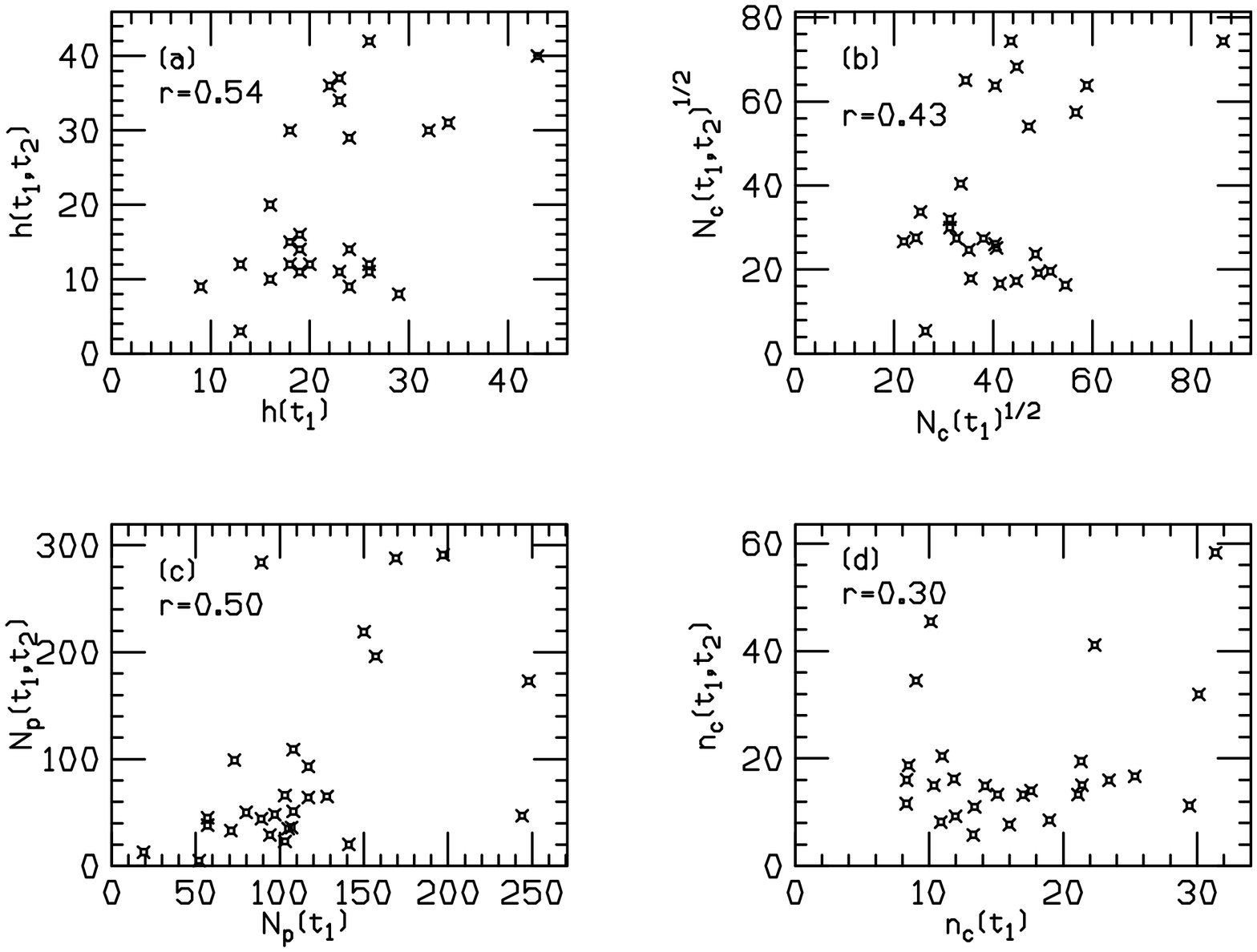}}
\caption{Predictive power of each indicator at year 1994 for the value of the same indicator for the papers published in the $1995-2006$ time interval
for the sample APS95.}
 \label{Fig. 2}
\end{figure}

It is easy to understand why the correlations here are weaker than in the first example . Scientists are elected to APS fellowship at very different
stages in their career, so the horizontal axis variables in these figures are not time-normalized. For example, a member of this group may have had
a large $N_c$ in 1994 because he/she had been publishing for many years at a slow rate, and his/her productivity in the subsequent 12 years would not be expected
to be larger than that of another scientist of this group that started his/her career much later and had a higher publication rate. 

Note also that the 12-year productivity and impact of the APS fellows sample (table II) is on average substantially higher than that of the random sample
PRB80 (table I), and that there are no points on the x-axis in the figures for the period $t_1-t_2$ for the APS sample (figs. 7,8) in contrast to that
of the PRB80 sample (figs. 4-6). These differences are to  be expected from the fact that election to APS fellowship is not a random process.

  \begin{figure}
\resizebox{9.0cm}{!}{\includegraphics[width=8cm]{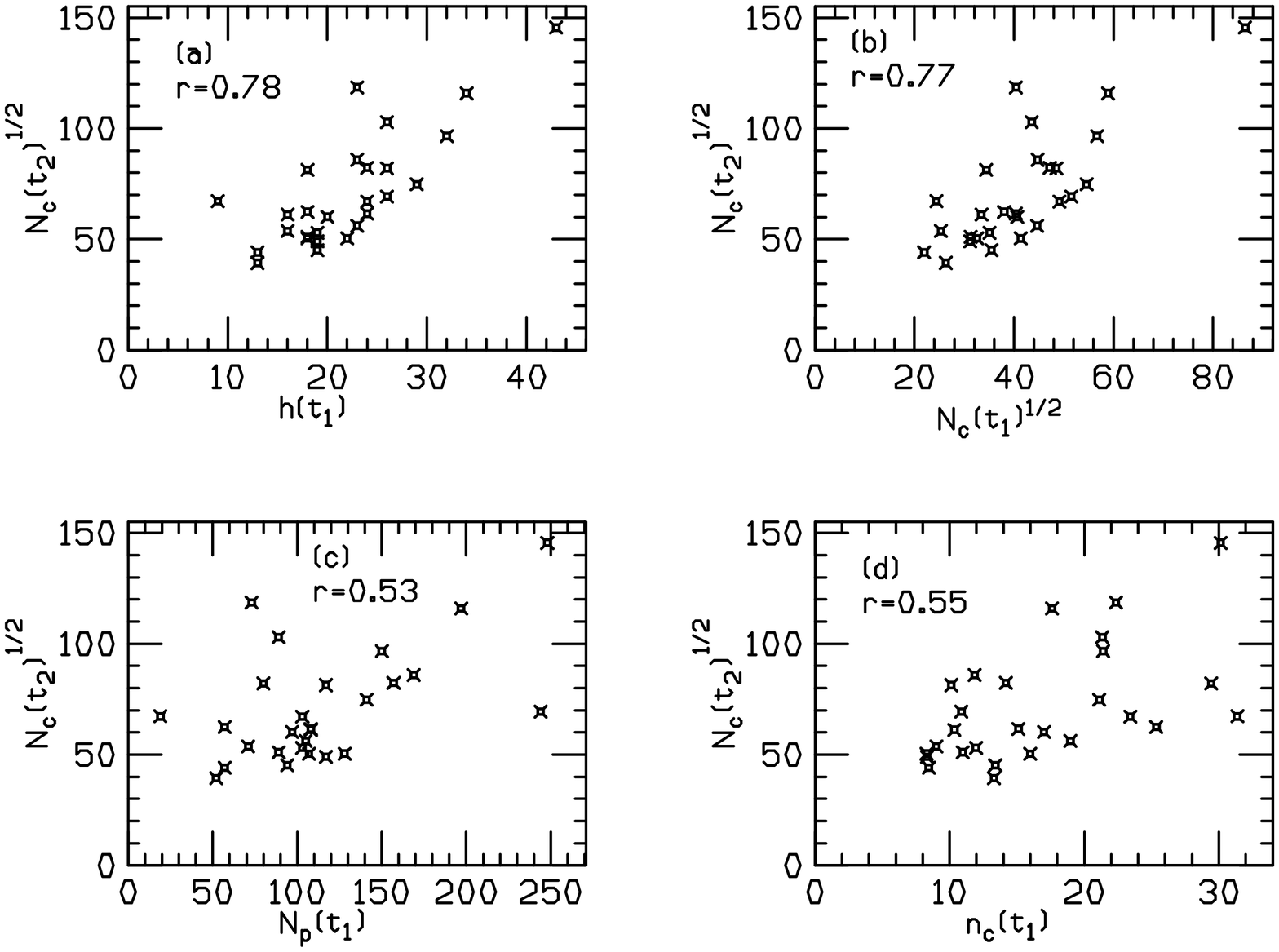}}
\caption{Predictive power of each indicator at year 1994 for the number of citations to  all papers published up to 2006 for the sample APS95.}
 \label{Fig. 2}
\end{figure}
 
  \begin{figure}
\resizebox{9.0cm}{!}{\includegraphics[width=8cm]{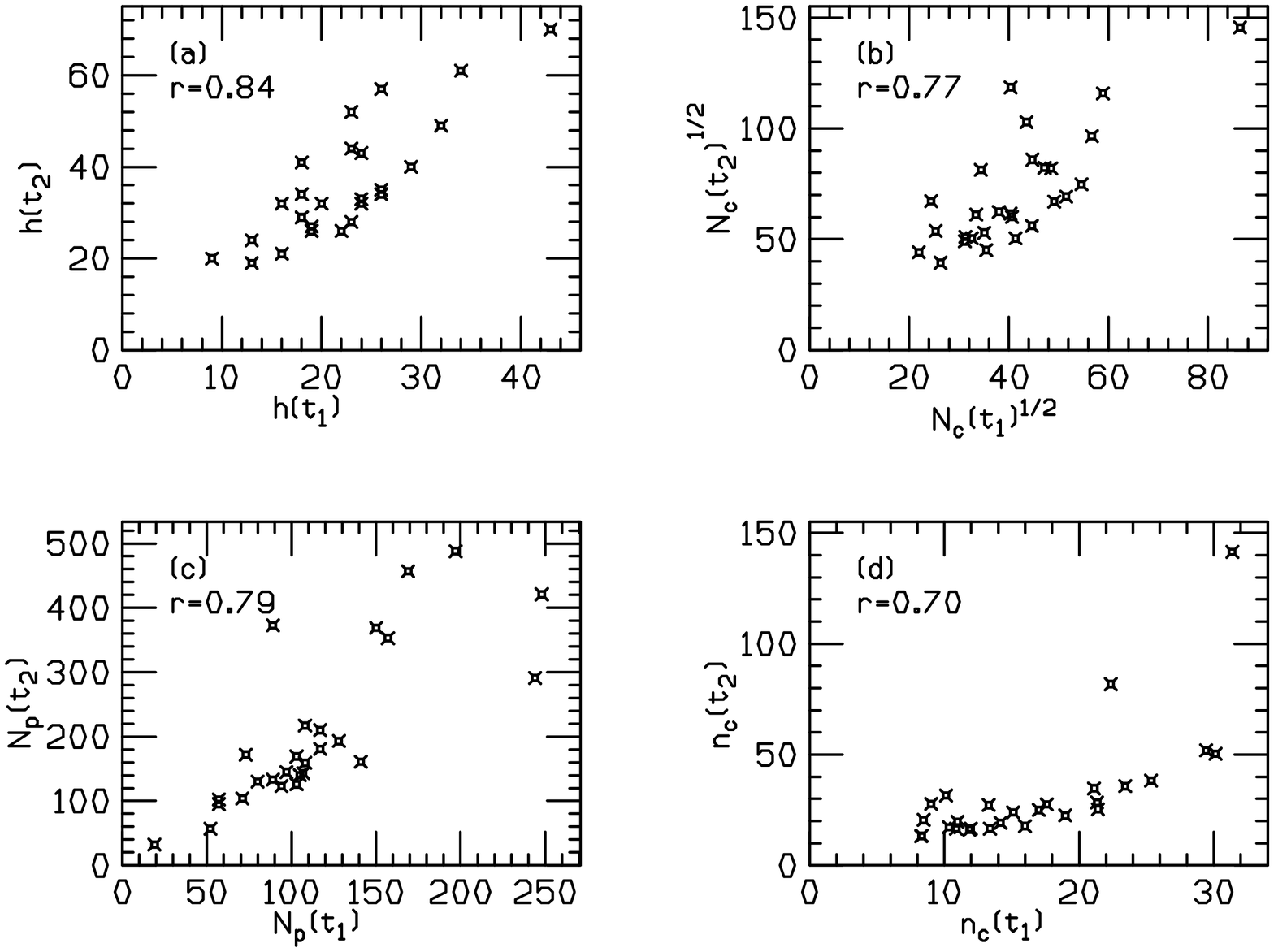}}
\caption{Predictive power of each indicator at year 1994 for the value of the same indicator at year 2006
for the sample APS95.}
 \label{Fig. 2}
\end{figure}
 
 \section{combining $h$ and $N_c$}
    Our results indicate that the h-index and the total number of citations are better than the number of papers and the mean citations per paper to predict future achievement, 
with achievement defined by either the indicator itself or the total citation count $N_c$. Furthermore, we found a small consistent advantage of the h-index compared
to $N_c$. 

It has been argued in the literature that one drawback of the h-index is that it does not give enough 'credit' to very highly cited papers, and various modifications have
been proposed to correct this, in particular Egghe's g-index\cite{egghe}, Jin's AR index\cite{jin} and Komulski's $H^{(2)}$ index\cite{kom}. These modified indices reward authors with higher citation numbers
in the papers that contribute to the h-count.

To test the possibility that giving a higher weight to highly cited papers may enhance the predictive power of the h-index, we considered the following 
expression:
\beq
h_\alpha \equiv \sqrt{h^2 +\alpha N_c}
\eeq
and asked the question: which value of $\alpha$ will result in $h_\alpha(t_1)$ best predicting the citation count of future work, $N_c(t_1,t_2)$? 
That is, we considered the cases of Fig. 4(a) and Fig. 7(a) with $h_\alpha(t_1)$ in the abscissa instead of $h(t_1)$.

The resulting correlation coefficients as function of $\alpha$ are shown in Fig. 11. Surprisingly, a small $negative$ $\alpha$ ($\alpha \sim -0.1$) yields the largest
correlation coefficient in both samples considered. For   positive $\alpha$, the correlation coefficients decrease monotonically and approach the
values corresponding to  the predictor $N_c$, $r=0.53$ and $r=0.43$ respectively, corresponding to Figs. 4(b) and 7(b).

  \begin{figure}
\resizebox{9.0cm}{!}{\includegraphics[width=8cm]{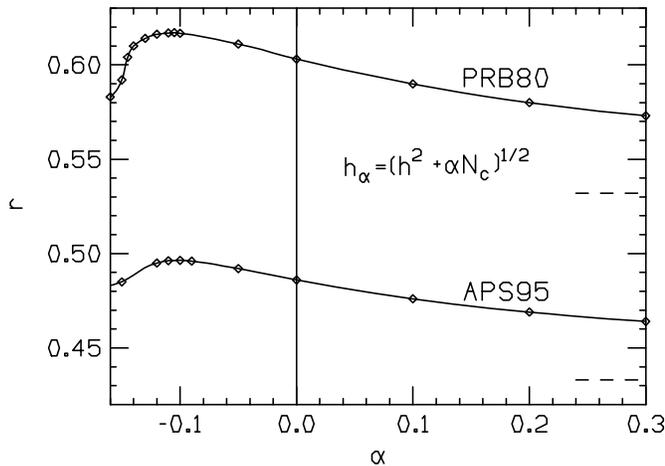}}
\caption{Correlation coefficient $r$ between $N_c(t_1,t_2)$ and $h_\alpha$ defined in Eq. (2) for samples PRB80 and APS95. As $\alpha$ increases, the curves approach
the asymptotic values given by the dashed lines, $r=0.53$ and $r=0.43$ respectively}
 \label{Fig. 2}
\end{figure}
 
 Consequently the best predictor of future achievement (achievement defined as number of citations) inferred from our data (e.g. sample PRB80) would be a linear regression fit to
 $\sqrt{N_c(t_1,t_2)}$ versus $h_\alpha(t_1)$ with $\alpha=-0.1$ (correlation coefficient $r=0.62$), leading to the paradoxical result that given two researchers with
 the same h-index, the one with $lower$ $N_c(t_1)$ should be expected to earn $higher$ number of citations   in the subsequent time period.

By using the relation $N_c=ah^2$ we can rewrite Eq. (1) as
\beq
h_\alpha = h \sqrt{1 +\alpha a}
\eeq
The fact that a negative $\alpha$ yields larger predictive power indicates that authors with large values of $a=N_c/h^2$ are on average less likely to earn a larger
number of citations in future work than authors with smaller $a$. We believe that this effect is principally due to the effect of coauthorship discussed in the next section.

\section{discussion}

In summary, we found that the h-index appears to be better able to predict future achievement than the other three indicators, 
number of citations, number of papers and  mean citations per paper, 
with achievement defined by either the indicator itself or the total citation count $N_c$.  In addition, the h-index was found to be a better predictor of productivity ($N_p$) than $N_p$ itself. Furthermore, in attempting to combine $h$ with $N_c$ to enhance the predictive power of $h$ we found that $N_c$ should enter with a $negative$ weight.

It is interesting and not obvious that the h-index is able to  predict both itself and the productivity $N_p$ better than $N_p$ can predict itself. Perhaps it indicates that some of the prolific authors with small citation counts feel less of an incentive to continue being prolific as they perceive that
their work is not having impact.

We believe the superiority of h compared to $N_c$ as a predictor is due to the issue of coauthorship, touched on in ref. \cite{hindex} and in the introduction. Let us elaborate
on this further.

Consider a paper $j$ with $N_c^j$ citations coauthored by several scientists with different levels of seniority as well as of ability, each of which made
different contributions to this paper. If we are counting citations, each coauthor gets the same 'credit', i.e. adds $N_c^j$ citations to his/her total
citation count, independent of his/her individual contribution to this paper.

Instead, if we are considering h, this paper will or will not contribute to the i-th author's h-index, $h_i$, depending on whether $N_c^j>h_i$ or $N_c^j<h_i$.
If it contributes, that author only 'needs' $h_i$ of the $N_c^j$ citations to increase his/her h by one. So one may say that each author $i$ gets
'allocated' only $h_i$ of the $N_c^j$ citations. Junior, as well as less able authors, are likely to have a lower $h$ than senior and more able
authors, $and$ they are likely to have made a lesser  contribution to 
the paper\footnote{Of course it will often be the case that a junior coauthor will have performed most of the actual work
for the paper. Nevertheless, if the paper has senior coauthors and ended up with a large number of citations, it will often be the case that the
senior coauthor/s will have played the crucial role.}. Hence it is appropriate that they benefit from a 
smaller portion of the total $N_c^j$.

In other words, to 'first order', using h rather than $N_c$ as a measure of scientific achievement automatically reduces an important source of
distortion when  multiply coauthored papers are involved, by allocating a smaller portion of the credit to those authors that are likely to have
contributed less. The argument is not foolproof and exceptions undoubtedly will occur, but the 'injustice' done to powerful junior coauthors
in the early stages of their career will  automatically be remedied in due time as their h-index rapidly increases.

Furthermore it is interesting and revealing that the advantage of $h$ over $N_c$ in predicting future $N_c$ values is lost when $cumulative$ rather than only
citations to new papers are considered (Fig. 2 vs. Fig. 4, Fig. 9 vs. Fig. 7). We suggest that this also reflects the effect of coauthorship. Highly cited papers
in the initial period will usually continue to garner a high number of citations in the subsequent period also for those among the  coauthors that 
made only minor contributions to the paper. While the cumulative citations of those individuals will be high, they should be less likely to make major $new$
contributions in the subsequent period. Thus we argue that even for a decision focused on
optimizing expected $cumulative$ achievement $h$ should be favored as an indicator as it appears better able to 
predict $individual$ cumulative achievement.

Other recently proposed bibliometric measures that give more weight to very highly cited papers such as  Egghe's g-index\cite{egghe},  
Jin's AR index\cite{jin} and Komulski's $H^{2}$ index\cite{kom} are likely to suffer from the same drawback as $N_c$, since they
 will assign the citations of highly cited papers equally to all coauthors without discrimination. 
Thus we conjecture that the predictive power of these modified indices  is likely to be worse than that of the h-index, as our analysis of the $h_\alpha$ index (Eq. 2) 
also suggests.

With respect to the indicator $n_c$, mean number of citations per paper, our results indicate that it has very little
predictive value. The low correlation found 
between $n_c's$ in the different timeframes (initial 12 years and subsequent 12 years) is due to a variety of reasons. In some cases the individual's productivity $N_p$ 
remained similar but the total impact $N_c$ changed substantially, sometimes both  productivity declined and the total impact declined even more, sometimes productivity increased and
the mean impact per paper also increased.

These results are at odds with the conclusions of the recent study by Lehmann, Jackson 
and Lautrup \cite{lehmann1,lehmann2}. They   start from the reasonable assumption that "the quality of a scientist is a function of his or her
full citation record" and address the question of which single-number indicator is best to discriminate between scientists,
aiming "to assign some measure of quality". 
They argue that an indicator is of no practical use unless "the uncertainty in assigning it to individual scientists is small". They perform a 
Bayesian analysis of citation data from a large sample extracted from  the SPIRES database\cite{spires}, and find that among the three indicators 
(i) mean number of citations per paper ($n_c$), 
(ii) number of papers published per year ($N_p/n$), and (iii) $h$-index, $n_c$ is far superior in discriminating between scientists.
They conclude   that  
{\it [c]ompared with the h-index, the mean number of citations per paper is a superior indicator of scientific quality, in terms of both accuracy and precision}, and hence that
"the mean or median citation counts (per paper) can be a useful factor in the academic appointment process".

Bornmann and Daniel\cite{bornmann}  echo their conclusions and state that  the Lehmann et al study
 "raises some doubt as to the accuracy of the (h-)index for measuring
scientific performance", that "the mean, median, and maximum numbers of citations are reliable and permit accurate measures of scientific performance", and instead that  "the h index is shown to lack the necessary accuracy and precision to be useful".

We argue that these conclusions are deeply flawed. Our results here have shown that the h-index is a far better predictor of future scientific
achievement than the mean number of citations per paper, and surely the same would hold for the median. For example, the correlation coefficient 
between the number of citations in the subsequent
12 years and the h-index in the previous 12 years in sample PRB80 was found to be $r=0.60$, much larger than  $r=0.21$, the correlation found with the mean number of citations per paper in the previous 12 years. The h-index was also far superior at predicting itself, $r=0.61$ versus $r=0.23$ for $n_c$. Similar pattern was found in our other sample, and 
it is likely that  similar results would be obtained with the sample used by Lehmann et al.

This example illustrates the 
danger in using sophisticated mathematical analysis to jump to practical conclusions (of sometimes life-changing consequences) in the delicate issues under considerations.
While the Lehmann et al study may be  correct in concluding that the mean number of citations is better to 'discriminate' between scientists for a given fixed time period according
to their definition, the fallacy in their argument appears to be that this does not imply that the indicator is associated with an identifiable individual trait that would be expected to persist with time. And certainly not with 'scientific quality'. Else,  one is forced to
conclude in light of the results of the present paper that 'scientific quality' (as defined by Lehmann et al)
 in the past is nearly uncorrelated with scientific quality in the future for
individual scientists, a conclusion that defies common sense.

Instead, a variety of studies\cite{hindex},\cite{bornmann2},\cite{vanraan},\cite{list}, have shown that the h-index by and large agrees with other
objective and subjective measures of scientific quality  in a variety of different disciplines\cite{h1,h2,h3,h4,h5,h6}, and the present study shows that the h-index is also effective in discriminating among scientists that will perform well and less well in the future. We conclude tentatively (assuming that future empirical studies
will corroborate the results of this paper) that the h-index is a useful indicator of scientific quality that can be
profitably used  (together with other criteria) to assist
in academic appointment processes and to allocate research resources.

 \acknowledgements 
The author is grateful to Marie McVeigh for helpful advice on  extracting information from the ISI Web of Science database,  to P. Ball for calling Ref. \cite{lehmann1} to his attention,
and to M. Cardona for stimulating discussions.


\begin{references}
\bibitem{hindex} J.E. Hirsch, Proc. Nat. Acad. Sciences (USA) {\bf 102}, 16569 (2005).
\bibitem{lehmann1} S. Lehmann, A.D. Jackson and B.E. Lautrup, http://arxiv.org/abs/physics/0701311 (2007).
\bibitem{lehmann2} S. Lehmann, A.D. Jackson and B.E. Lautrup, Nature {\bf 444}, 1003 (2006).
\bibitem{egghe} L. Egghe, Scientometrics {\bf 69}, 131 (2006).
\bibitem{jin} B.H. Jin et al, Chinese Science Bulletin {\bf 52}, 855 (2007).
\bibitem{kom} M. Komulski, ISSI Newsletter {\bf 2}, 4 (2006).
\bibitem{spires} http://www.slac.stanford.edu/spires/hep/
\bibitem{bornmann} L. Bornmann and H.D. Daniel, Jour. of the Am. Soc. for Inf. Sci. and Tech. {\bf 58}, 1381 (2007).
\bibitem{bornmann2} L. Bornmann and H.D. Daniel, Scientometrics {\bf 65}, 391 (2005).
\bibitem{vanraan} F.J. Van Raan, Scientometrics  {\bf 67}, 491 (2006).
\bibitem{list} See extensive list of references in ref. \cite{bornmann}.
\bibitem{h1}   C.D. Kelly and M. D. Jennions, Trends Ecol. Evol. {\bf 21}, 167 (2006).
\bibitem{h2}  K.T.  Jeang, Retrovirology {\bf 4}, 42 (2007).
\bibitem{h3} B. Cronin and L.I. Meho, Jour. Am. Soc.  Inf. Sc. and Tech. {\bf 57}, 1275 (2006).
\bibitem{h4} J.E. Iglesias and C. Pecharroman, arXiv:physics/0607224 (2006).
\bibitem{h5} C. Oppenheim,  Jour. Am. Soc.  Inf. Sc. and Tech. {\bf 58}, 297 (2007).
\bibitem{h6} Chemistry World Vol. 4, No. 5 (2007).


 \end{references}
\end{document}